\documentclass[10pt,letterpaper]{article}
\usepackage{epsfig}

\begin{document}

\title{\large \bf Networking in the Physical World\\}
\author{Michael Neufeld and Craig Partridge\\
BBN Technologies \\
10 Moulton St. \\
Cambridge, MA 02138 \\
mneufeld@bbn.com, craig@bbn.com
}

\maketitle

\subsection*{Abstract}
In this work we propose a network {\em meta-architecture} based on
fundamental laws of physics and a physical model of computation.
This meta-architecture may be used to frame discussions about
novel network architectures as well as cross-layer alterations
to the canonical network stack.

\section{Introduction}
\label{sec:intro}
The canonical layered network stack architecture has been leveraged
successfully to create a world-wide computer network.
However, technological advances have exposed limitations and
weaknesses of the layered architecture.
Large quantities of inexpensive computing power has encouraged
software to encroach on the previously hardware-only physical layer.
Quantum computing and quantum information theory have introduced new
types of communication channels that do not necessarily fit cleanly into
the existing stack architecture\cite{bbnquantum}.
Much discussion has arisen around how to safely introduce cross-layer
mechanisms into the stack\cite{kawadia}, replace layers in the stack
that have proven problematic in emerging applications, and
even replacing the stack with something else entirely\cite{braden03}.
Unfortunately, most discussion surrounding cross-layer systems lacks
a coherent and comprehensive set of goals: cross-layer dependencies
are simply introduced as required to work around barriers in the
existing network stack. Many ``clean slate'' networking
architectures suffer from an excessively abstract
point of view that is little help in actually constructing 
physically realizable and scalable systems. Conversely,
approaches that make ad hoc modifications to the existing stack
in the interest of functionality may not produce any real insight
into the broader nature of networking systems.
In this work we seek to rectify this situation by proposing a
network {\em meta-architecture} that may be used to analyze
and frame discussion about the canonical network stack and
cross-layer modifications as well as clean slate architectures
for the future. Our approach will begin with a model for computation,
storage, and networking that is grounded in the laws of physics as
we currently know them. On top of this foundation we will then proceed
to explore issues that are of importance to networking and storage systems.

\section{A Physical Foundation for Computation}
\label{section:physicalfoundation}
We will now outline a ``bottom up'' approach to abstract symbolic
computation that directly relates
abstract logical computation and physical operations.
Instead of proposing a {\em symbol} manipulation machine
we will begin with the manipulation and semantics of {\em physical states}
and how to transform physical states into symbols. This model is based
on fundamental tenets of modern physics, and was constructed under the
influence of work by Feynman\cite{feynman}, Zuse\cite{zuse} and
Fredkin\cite{fredkin} regarding fundamental notions of computation
and physics. We do not claim that this is a novel model, and this
work is not primarily about computation. However, we feel that
networking, storage, and computation are all fundamentally related
tasks, and that having a firm notion of a computational model is
critical for contextualizing and understanding networking and storage.
Symbolic computation relies on physical entities. Abstract symbols
map to physical entities and symbolic computation
specifies changes in the physical states of those entities.
Storage and networking move symbols by moving entities or properties
of entities in time and space. Unless situated
in some model of space and time, storage and networking
are symbolically indistinguishable from the identity operator.
Even systems that strive to eliminate physical
reality introduce {\em some} model of location and time, {\em e.g.}
memory addresses correspond to physical locations in a memory store, and
the sequential execution of instructions marks the passing of time.
IP addresses correspond to locations as well as names of systems,
though this conflation of name and location causes well-documented
problems when entities are allowed to move\cite{mobileip:rfc}.
Counters used by logical clock systems express temporal event ordering
and causality without insisting on non-local clock synchronization.
Systems and applications operating directly in real space and time
often use much more realistic models of space and time, {\em e.g.} the
Global Positioning System (GPS) must account for relativistic and
rotational motion effects on time and photons to obtain precise
location estimates.
We will refer to a physical entity as an {\em artifact}. An artifact
has properties of physical particles or waves:
location, mass, energy, momentum, charge, {\em etc.}, and quantum
and relativistic effects may be important.
An artifact may be a fundamental object, or assembled from other artifacts.
Compound artifacts often behave differently from their subcomponents,
introducing new levels of abstraction which are critical for managing
complexity. For example,
chemical analysis is facilitated by treating subatomic particles 
as aggregate atoms and molecules, and biology is greatly assisted
by using cells as a basic abstraction.

An artifact may have attributes that correspond to some
abstract unit of information, or {\em symbol}.
When an artifact has such symbolic attributes
we refer to that artifact as a {\em glyph}.
Artifacts may interact with other artifacts. When an artifact
alters another artifact we refer to the alterer as a {\em scribe}.
Artifacts may be both glyphs and scribes; in fact
a glyphs' action as a scribe is often used to observe it.
{\em Physical operations} that alter
artifacts may correspond to {\em symbolic operations}.
A glyph must be {\em measured} to extract its corresponding symbol.
This measurement is important and the subject of much engineering in classical
computation systems. In quantum computation measurement, whether intentional
or unintentional, is absolutely vital because the state superpositions upon
which quantum computations rely collapse when observed.
The measurement of a physical state
includes some noise and uncertainty, ultimately limiting the number of symbols
that may be reliably discerned from a particular physical state.
Shannon's bit has emerged as the predominant way of 
handling noisy symbol streams, providing mathematical guarantees
on the number of coded symbols that may be reliably embedded in
a classical property of a glyph. Glyph systems used in digital
computation systems may also be inherently quantized, providing
an easy transition into bits so long as noise levels remain sufficiently low.

The basic mapping between glyphs and corresponding symbols is defined by the
denotational semantics\cite{denotationalsemantics} of the information system,
and additional layers of symbolic meaning may be defined by further
denotations.
We will not delve into a philosophical discussion of what the symbols
{\em themselves} mean, or if it is meaningful to discuss the notion of a
symbol outside of some physical embodiment.
Instead we will simply state that the symbols may have
further symbolic denotations defining their meanings and leave the rest to
traditional denotational semantic techniques. Our only requirement
is that any abstract symbol must eventually correspond to some physical
embodiment.
The operational and axiomatic constraints of glyphs are defined
by the physical laws governing the glyphs, and these glyph semantics
will ``bubble up'' and influence feasible symbolic
operational\cite{operationalsemantics} and
axiomatic\cite{axiomaticsemantics} semantics.

\begin{figure}
\centerline{
\begin{tabular}{ccc}
\epsfig{file=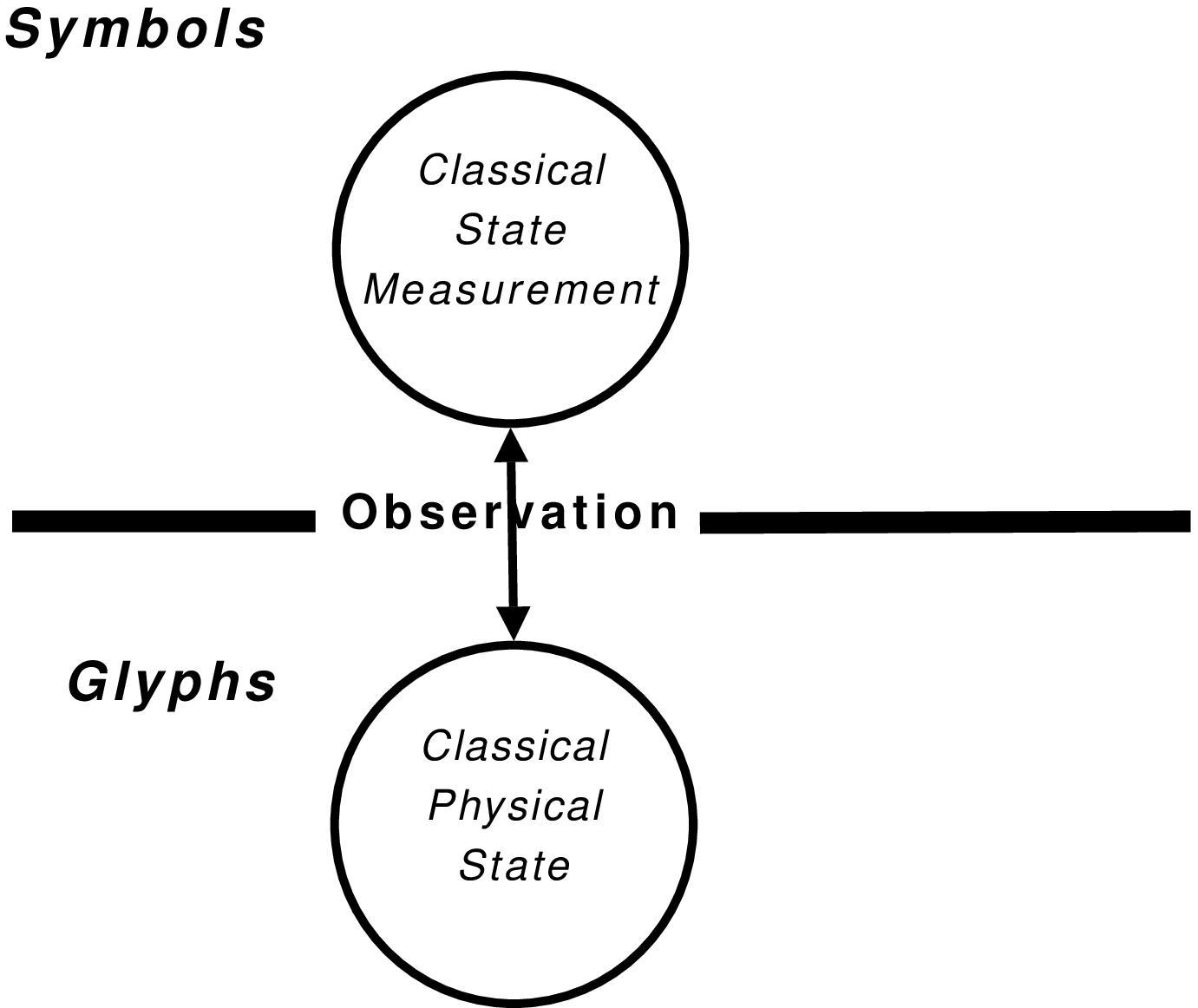,width=0.3\columnwidth}
&
\epsfig{file=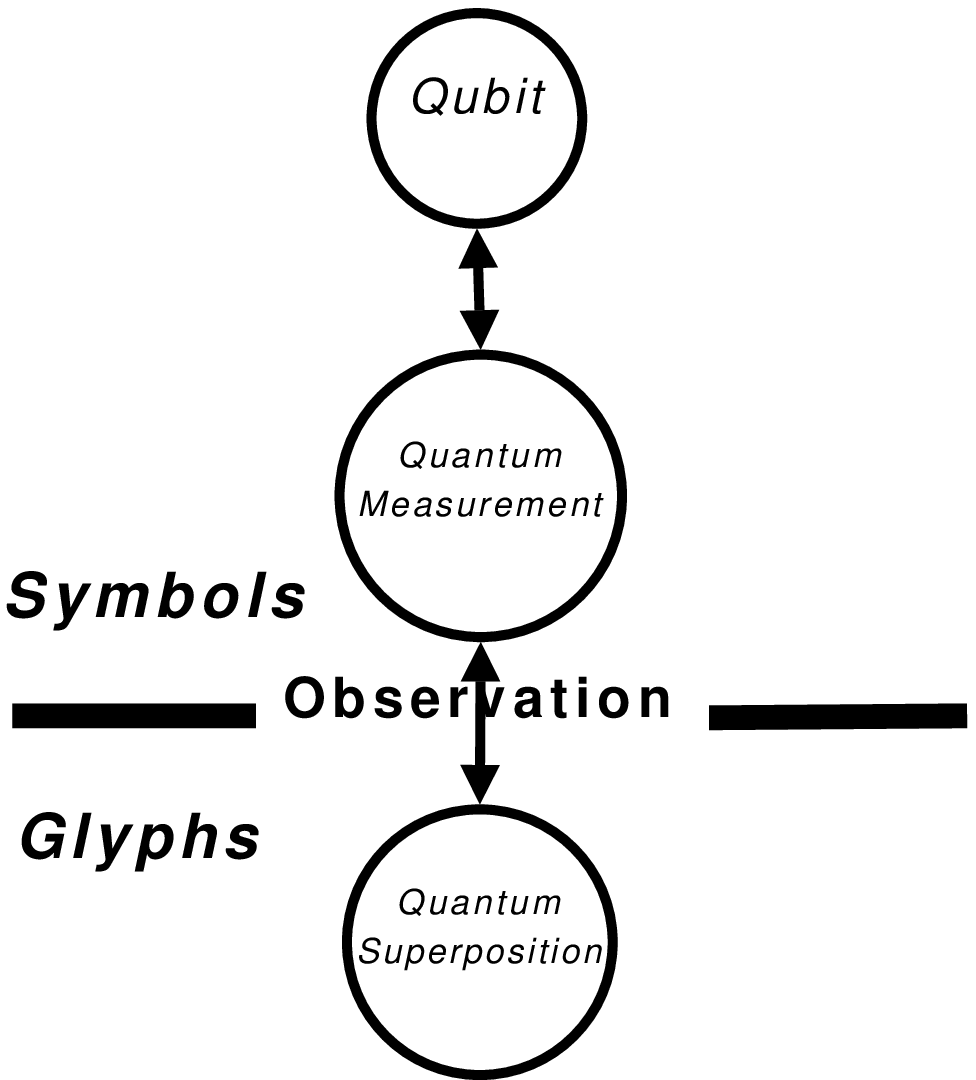,width=0.3\columnwidth}
&
\epsfig{file=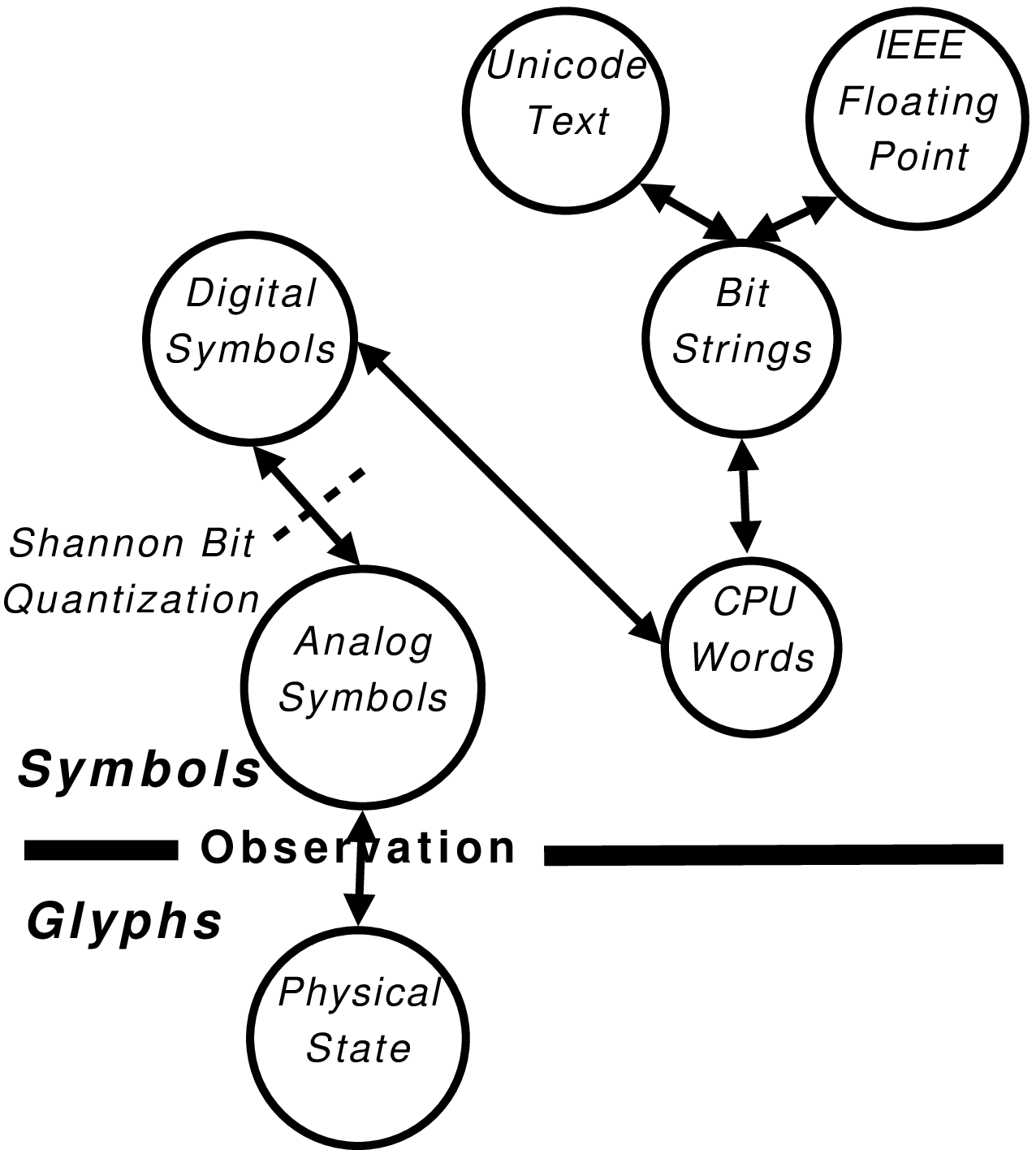,width=0.3\columnwidth}
\\
\begin{minipage}[c]{0.3\columnwidth}
\centerline{
\footnotesize Analog Computation
}
\end{minipage}
&
\begin{minipage}[c]{0.3\columnwidth}
\centerline{
\footnotesize Quantum Computation
}
\end{minipage}
&
\begin{minipage}[c]{0.3\columnwidth}
\centerline{
\footnotesize Digital Computation
}
\end{minipage}
\end{tabular}
}
\caption{\footnotesize A diagram showing how glyphs, symbols, and filtering combine to create symbol sets for a analog, quantum, and traditional digital computation systems.}
\label{figure:symbol_systems}
\end{figure}
Figure~\ref{figure:symbol_systems} shows diagrams of how
symbols for analog, quantum, and digital computation system are formed from
glyphs. In the digital system, glyphs are measured to create analog symbols,
which are then further processed to generate digital symbols, and then mapped
into bit strings of machine word size. These machine words may be further 
mapped into operators of a microprocessor instruction set, Unicode
characters of human language, or any number of other symbolic domains.
Unfortunately the term ``bit'' is hopelessly overloaded: a ``bit'' not
only refers to a specific way of converting a noisy signal into
discrete symbols, but also the use of binary digits as a universal
symbolic alphabet. When working with physically discretized systems
it is convenient to use symbolic bits to represent states, 
but this is separate from performing bit-style signal filtering and
discretizing.
Current digital systems use multiple
glyph repositories for different tasks. CPU registers are used for
rapid, deterministic symbol manipulation embodied in a small number
of glyphs, cache memory near the CPU registers stores a larger number
of glyphs for rapid copying into registers, and external
RAM and disk-based swap farther away holding an even greater number of
glyphs. Typically all of these glyph stores are managed as a single
virtual space that is accessed via sequential labels.
Different glyph and symbol systems may have different sizes, potentially
requiring padding or {\em fragmentation/reassembly} to match a symbolic
system to a particular type of glyph.
Quantum computing systems operate primarily on pure glyphs, making
a symbolic observation only at points when collapsing superimposed
quantum states has a very high chance of producing the desired computational
result.

\section{Physical System Requirements}
Choosing an embodiment for networking, storage, or computation
is fundamentally motivated by physical system characteristics.
\begin{figure}
\centerline{
\begin{tabular}{cc}
\epsfig{file=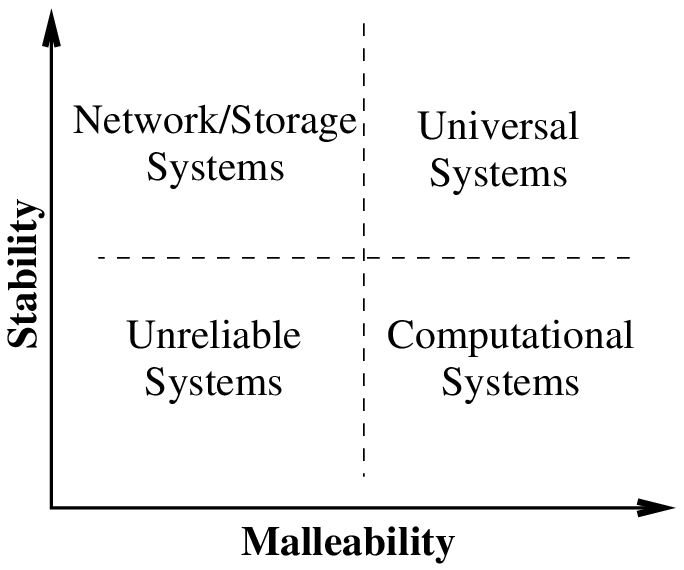,width=0.4\columnwidth}
&
\epsfig{file=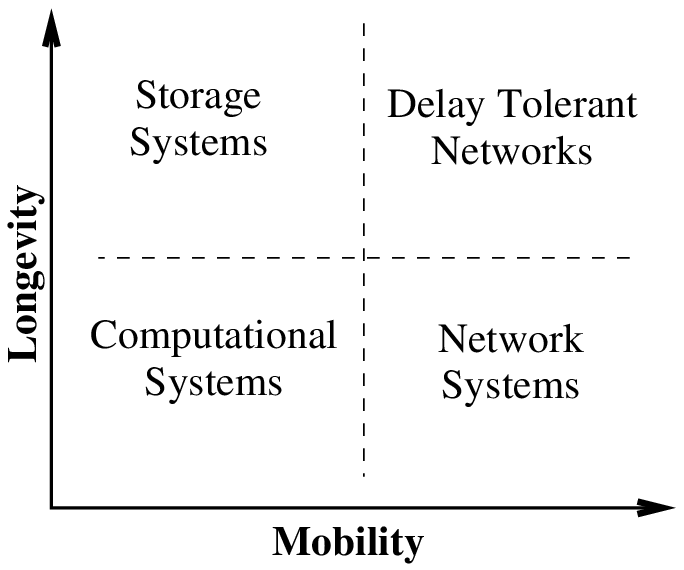,width=0.4\columnwidth}
\\
\begin{minipage}[c]{0.45\columnwidth}
\centerline{
\footnotesize Operational Stability and Malleability
}
\end{minipage}
&
\begin{minipage}[c]{0.45\columnwidth}
\centerline{
\footnotesize Mobility and Longevity
}
\end{minipage}
\end{tabular}
}
\caption{\footnotesize Computational, networking and storage systems place different demands on glyphs. Stability and malleability measure the difficulty and speed of glyph alteration respectively. Mobility and longevity measure how difficult it is to move the glyph in space and time respectively.}
\label{figure:compstorexfer}
\end{figure}
In traditional computation
systems it is typical to use fast-moving bosons or a media-based waves for
networking, and easily-trapped fermions for computation and storage.
However, this choice is not universally true.
As a microscopic example, consider the general operation of a cell. The normal 
operational ``computation'' is carried out largely by proteins.
These proteins are synthesized by cellular machinery driven by
RNA, which in turn is synthesized from DNA stored in the nucleus.
Tightly wound DNA is very compact and stable, making it an excellent
medium for storage, but cannot be used directly by cell machinery to
synthesize proteins. Transcribing segments of DNA into a more reactive
and expanded form (RNA) allows the cell to synthesize the amino acids
and proteins that drive cellular operation.
As further examples, chemical and molecular
communication schemes have been proposed for nanomachines\cite{molecular},
and the {\em Voyager} space probes contain discs intended to communicate
information to any extraterrestrial entities that might stumble
across them\cite{voyagerwebsite}, likely at a very great distance
in space and time.
From a general system architecture design standpoint
we propose four critical physical properties: {\em stability}, {\em i.e.}
the difficulty/energy required to change a state, {\em malleability},
{\em i.e.} how quickly it may be changed, {\em longevity}, {\em i.e.}
how easily it may be moved through time, and {\em mobility}, 
{\em i.e.} how easily it may be moved through space. These properties are
illustrated in Figure~\ref{figure:compstorexfer}. Expressing these
properties quantitatively relies on an ability to handle physical space,
time, and energy. Because networking and storage systems strive
to preserve symbolic state we are also interested in {\em distortion},
{\em i.e.} unintended changes in glyph and/or symbolic state. In the
next two sections we will discuss space, time, energy, and distortion
in greater detail.

\subsection{Space, Time, and Energy}
Space, time, and energy are three obvious pieces of physical information
that may be abstracted out of a physical system. In fact, these
physical properties emerge throughout networking systems as well
as physically-oriented applications, {\em e.g.} sensing and actuating
systems.
Current systems have varying notions
of time, particularly when accuracy and precision greater than that
specified by the POSIX {\em gettimeofday} call. Spatial location
is largely {\em ad hoc}, but a latitude and longitude are widely-used
as a coordinate system, largely
driven by the increasing availability of GPS receivers.
Naturally, latitude and longitude are not the end of the story: the
DARPA XNAV\cite{xnav} program proposes the use of x-ray pulsars for
determining location and time in deep space, and a nano-scale machine
would fare better with a correspondingly
small coordinate system. Furthermore, space and time are inextricably linked.
Most existing computational systems ignore relativistic factors, but
as clocks become more and more precise and operating environments more and
more extreme system designers will be forced to cope with modern physics,
{\em e.g.} the GPS system itself requires accounting for relativistic effects
in order to provide accurate location and time information.

\subsection{Distortion}
{\em Distortion} is perhaps not as obviously a fundamental physical
concern as space, time, and energy, but distortion is critical for
information systems.
In this work we define distortion as an observable deviation from
some ``normal'' state. Distortion is critical in networking and storage
systems because networking and storage strive to produce changes in
place and time {\em without} altering symbolic meaning, and distortions
in glyphs may result in distortions in symbols. The effects of distortions
have been extensively studied as noise in signal theory, Shannon's bit is
a technique for handling distortions ({\em i.e.} noise) in an analyzable
and generic fashion in quantized systems. Distortion generally only shows
up at the lower layers in the canonical network stack because of Shannon's
results regarding the separation of source and channel coding, but 
our meta-architecture does not require this split, {\em e.g.} for the 
exceptions to source/channel separation\cite{sensorcoding}.
A more simplistic reason to permit violation of source and channel
separation is the {\em cost} of performing encoding and decoding.
The time and energy required for encoding and decoding in an
information-theoretically optimal fashion may result in an overall
suboptimal or inadequate computational system, and generic networking
and storage systems should permit the computational portion of the
system to make such cost/benefit decisions.

\section{Names, Addresses, Locations, Times, and Routes}
\label{section:names}
Earlier work\cite{names-hauzeur} has stressed the fundamental nature
of names, addresses, and routes. We will
take a physical approach to names and addresses, starting with the notion
that location and time are inherent properties of
artifacts that may be observed and interpreted.
{\em Names} are symbols associated with artifacts based on properties
of those artifacts. 
We will refer to names that depend directly or indirectly on location
and/or time as {\em addresses}. An address that depends
exclusively on location and/or time may be used as part of a
generic coordinate system for describing locations and times:
we will refer to such names as {\em pure addresses}.
We call location-independent names {\em labels}.
Because we are treating location and time as universal artifact properties
a label, by defining artifact(s), may define a set of addresses, and conversely
addresses may define some set of labels.
This fundamental connection between addresses and names can lead
to confusion. For example, we consider an IP ``address'' to be 
a label because it is not defined by physical location.
IP labels may be correlated with terrestrial location, but this
is a side effect of how the labels have been allocated, {\em not} a
fundamental property of the labels themselves.
At a low level, addressing (coordinate) systems require 
reference points. An addressing scheme may have reference points
defined in terms of other addressing schemes, and for abstract systems
this process may continue indefinitely. However, for physical systems
the process must be bootstrapped by an addressing scheme that does not
depend on some other addressing scheme, {\em i.e.} in terms of some
fundamental rigid body that is identifiable without referring to location.
A {\em label} may be used to identify an origin that is external to a
computational device, but the location of that origin must still be
determined with respect to each computational device. We are left with
the idea that each computational device must start with the ability
to refer to ``self'' as a label and ``here'' as a location. The next
logical step is for a system to have a model of itself situated in space
and time that may then be used to interact with the world.
To this end, we propose that physical time and location as well as energy be
treated as core system abstractions, and that tracking time and location
as core system services..
Physical time and location may then be used
to build shared addressing systems and to map between labels and
addresses; the addition of energy permits stability, malleability,
longevity, and mobility to be expressed.
Multiple computational devices may define common addressing systems based on
shared labels and artifacts that may then be used to construct
shared maps of computational artifacts. Furthermore, physical time
and location may be used for sensing 

\begin{figure}
\centerline{
\begin{tabular}{cc}
\epsfig{file=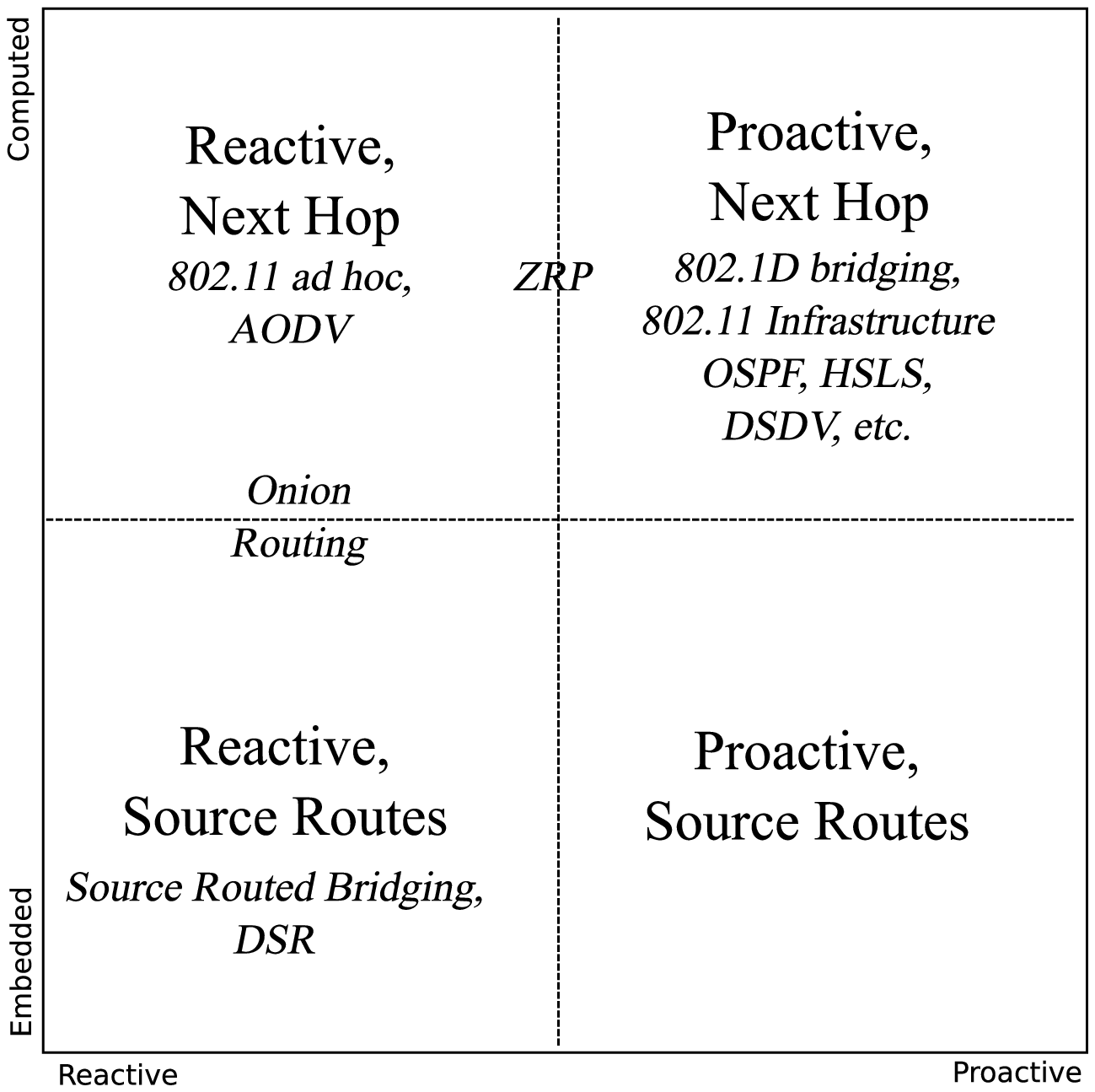,width=0.45\columnwidth}
&
\epsfig{file=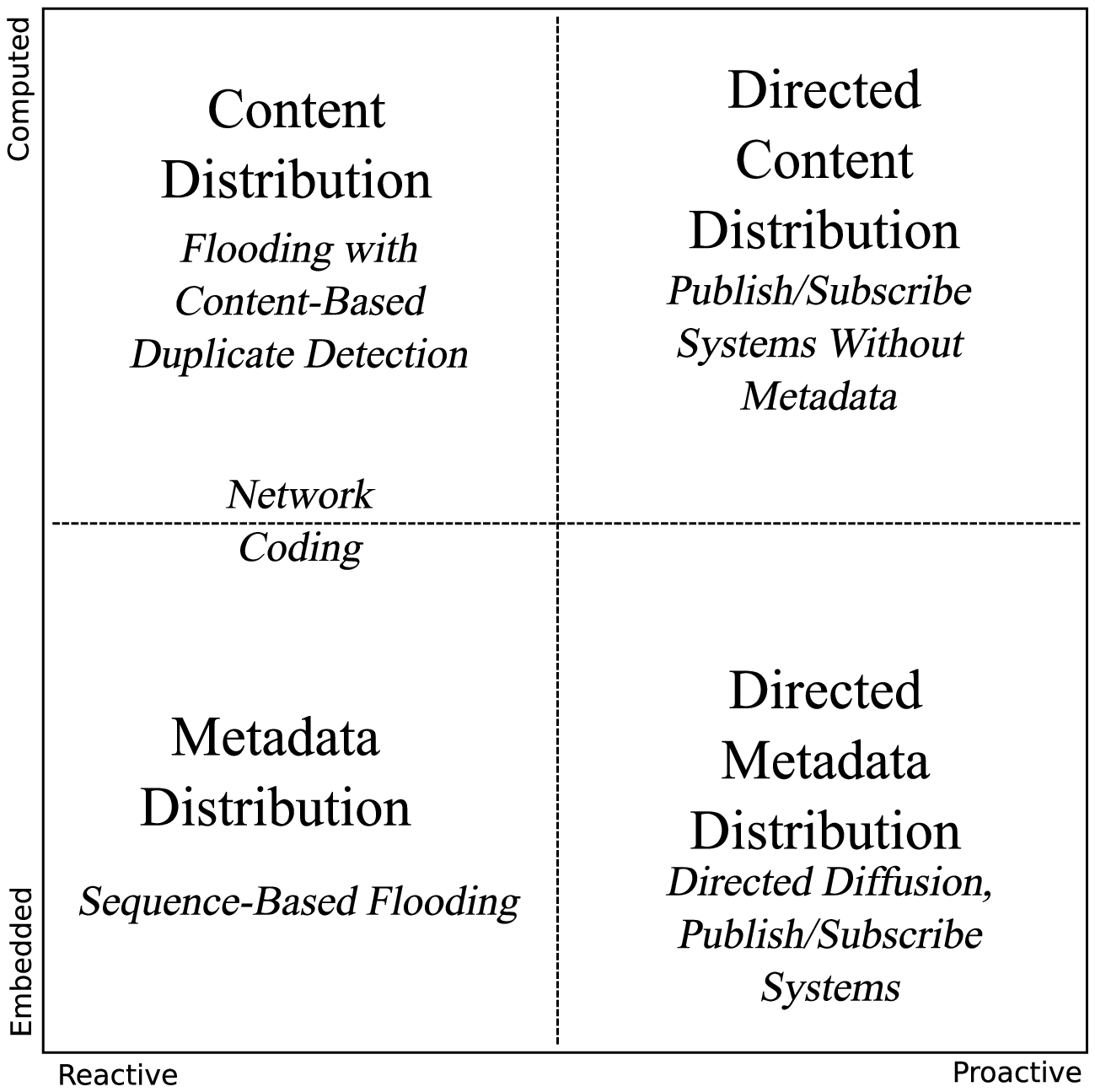,width=0.45\columnwidth}
\\
\begin{minipage}[c]{0.45\columnwidth}
\centerline{
\footnotesize Delivery Based on Endpoints
}
\end{minipage}
&
\begin{minipage}[c]{0.45\columnwidth}
\centerline{
\footnotesize Delivery Based on Content
}
\end{minipage}
\end{tabular}
}
\caption{\footnotesize Route information may be embedded alongside payload or computed along the way. Routing decisions may be based on the content itself or on the endpoints to visit, and routes may be preemptively discovered and maintained or discovered and maintained on demand.}
\label{figure:routing_categories}
\end{figure}

Core problems for networking systems are maintaining shared world maps and
determining the routes and paths that are to be taken by glyphs.
We will now broadly categorize
different techniques that may be employed for routing.
Routes may be precomputed and {\em embedded} with glyphs, as with
source routes, or {\em computed} locally as glyphs travel, as with
next hop routing. Routes may be calculated {\em proactively} ahead
of time or {\em reactively} on request.
Routes may be determined by {\em endpoints} or
the symbols and glyphs ({\em content}). Endpoint routing is commonly used in
existing topology-based routing schemes. Content-based routing does not
track endpoint topology. Instead, glyph semantics
\cite{siena:routing} or cleverly merging symbols or glyphs\cite{netcoding}
may be used to manage the flow of glyphs.
Figure~\ref{figure:routing_categories} illustrates how a selection of existing
network routing protocols manage the motion of information between endpoints.
Our categorization captures somewhat different aspects of routing
than the categorization proposed by Hauzeur\cite{names-hauzeur}.
Our categorization does not capture the localization or frequency
of route update calculations, but rather is concerned with what
aspects of a system require labeling and addressing to function
and how data and metadata must be managed in a system. In most traditional
layered digital systems metadata is kept at the ``front edge'' of the
glyph and completely separate from the data. This arrangement of data
and metadata has proven convenient, but is by no means the only possible
arrangement. For example,
the {\em Perspecta}\cite{perspecta} movie
audio system contained sub-audible routing metadata for an audio stream
alongside the audible soundtrack itself, and
optical networks
may opt to use light frequency to encode routing information.
A ``heap'' structure for packet metadata has also been proposed
\cite{braden03}, though we are not aware of any deployed systems
built around this principle.
Chemical and molecular networking schemes proposed
for nanomachines\cite{molecular} may use elements of both content
and endpoint routing,
and in general we feel that some combination of techniques should
be supported to handle a broad range of environments.
A further consideration for routing is {\em flow control}. Realistically
there are limits on the number of glyphs that may occupy a single region
of space at a given time, and flow control ensures that glyphs do not
overfill containers. In the canonical network stack flow control
is performed by MAC layers at individual links and over
multiple links by higher layer protocols such as TCP. Applications
may also perform their own flow control based on how quickly they
can process and generate information.

\section{System Meta-Architecture}
\label{section:systemarchitecture}
\begin{figure}
\centerline{
\begin{tabular}{cc}
\epsfig{file=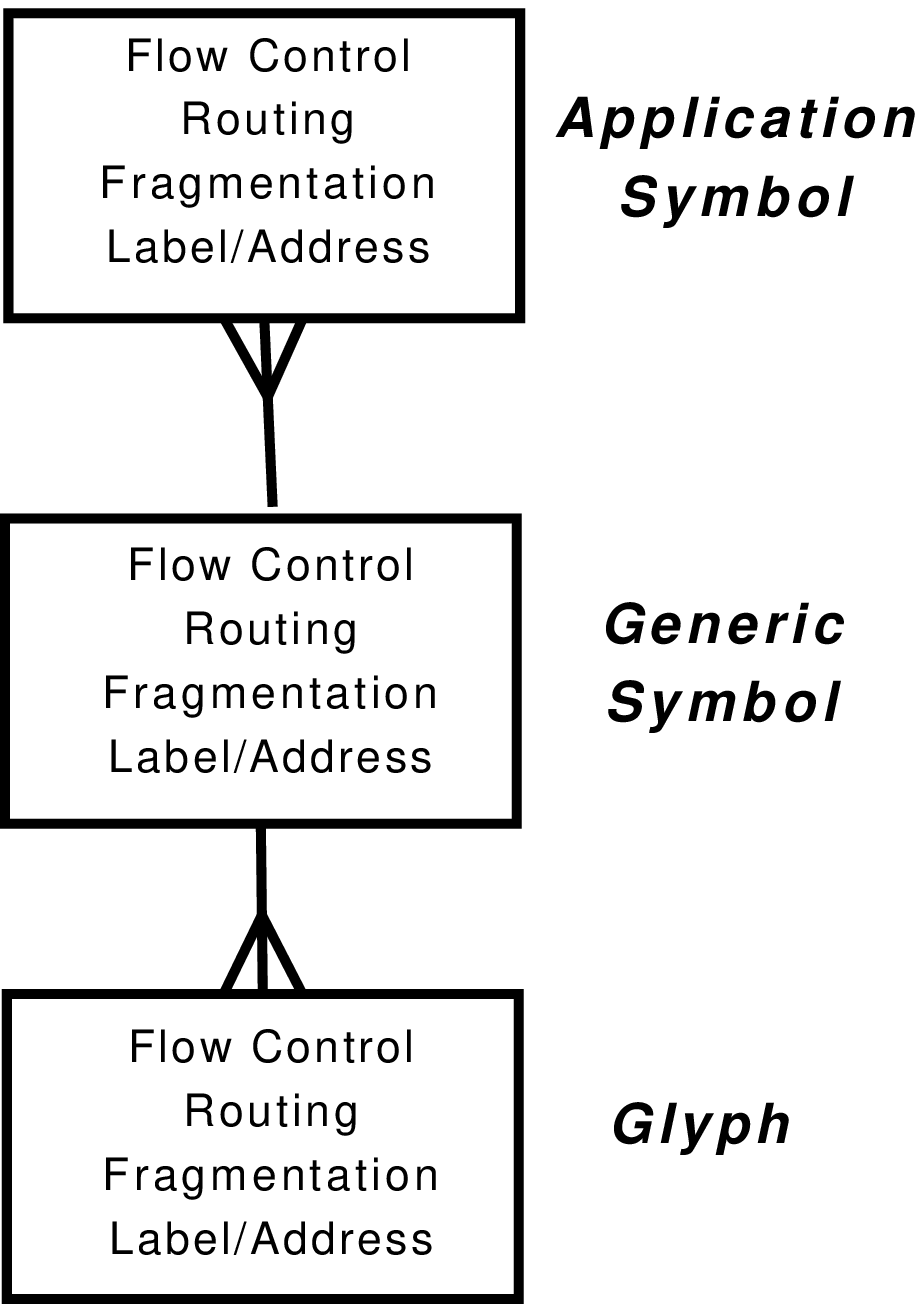,width=0.24\columnwidth}
&
\epsfig{file=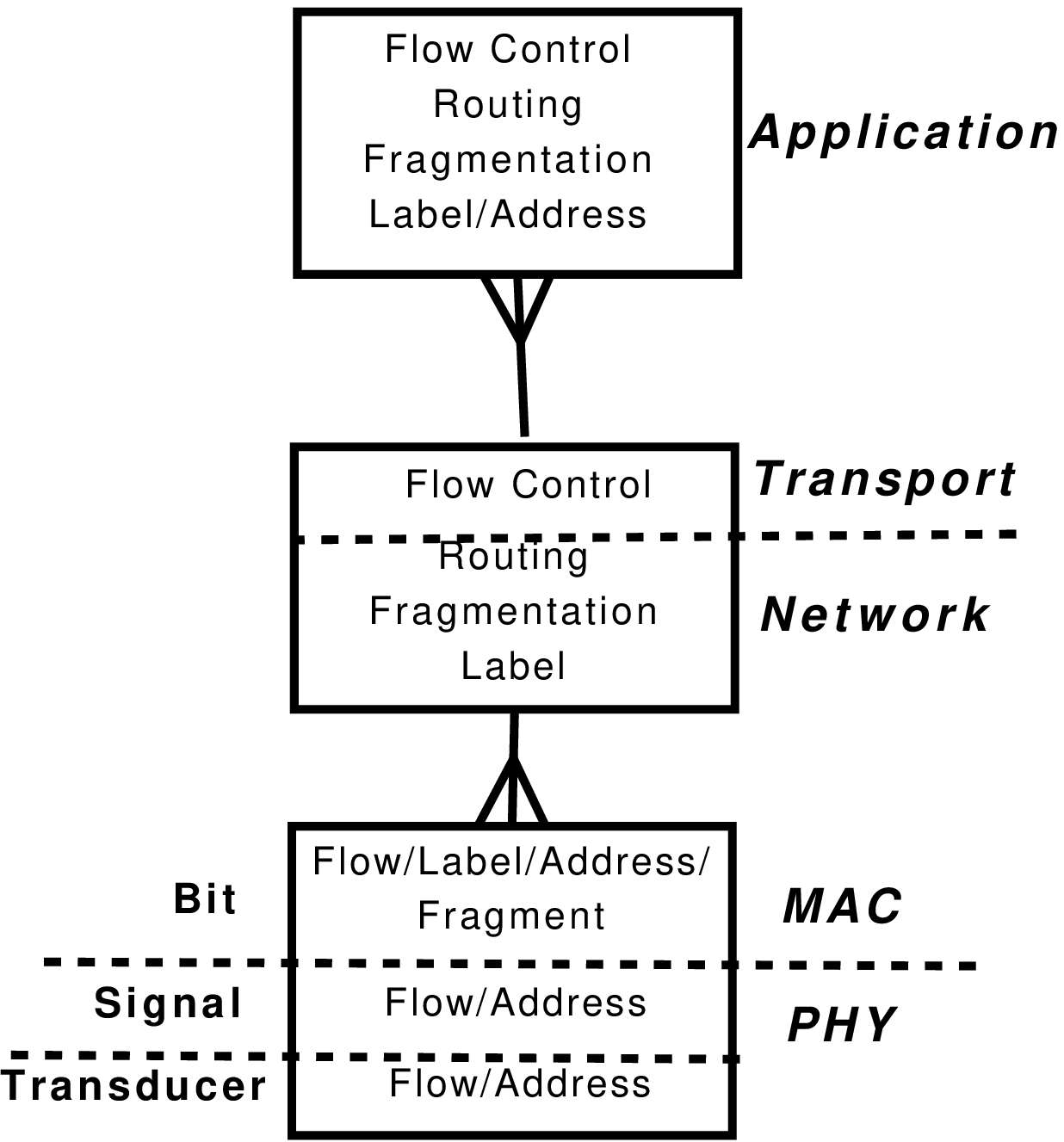,width=0.3\columnwidth}
\\
\begin{minipage}[c]{0.45\columnwidth}
\centerline{
\footnotesize Meta-Architecture
}
\end{minipage}
&
\begin{minipage}[c]{0.45\columnwidth}
\centerline{
\footnotesize Canonical Stack in Meta-Architecture
}
\end{minipage}
\end{tabular}
}
\caption{\footnotesize A picture of our proposed network meta-architecture. The ``crow's feet'' connectors imply a potential one-to-many relationship, and the inside of each box enumerates the networking functions performed in each layer. The canonical stack shown also illustrates how Shannon's bit may be used to provide another set of layers based on measured signals and signal processing, forming another ``waistline'' at the bit transition. The canonical Software Defined Radio architecture\cite{mitola95:sdrarch} is built around the Shannon waistline and permits the entire gamut of network functions to be performed at the glyph-aware layer.  Corresponding waistlines may also exist on the top end of the stack, such as the widespread use of XML as an interchange format for expressing application data semantics.}
\label{figure:system_meta_architecture}
\end{figure}

In this section we will outline a networking system meta-architecture. In
formulating this design we have drawn on a wealth of prior architectural 
systems work at the application layer\cite{structuredstreams},
internetworking layer\cite{nimrod:rfc,fara}, MAC and physical
layers\cite{bose98software,etx,ett}, and work spanning
layers\cite{alf,ramanathan,mitola95:sdrarch} for inspiration.
Our major meta-architectural goals are fairly straightforward to state:
we desire the ability to design system architectures that are {\em scalable},
{\em optimal}, {\em efficient}, {\em modular}, and {\em generic}.
Furthermore, these system architectures must be feasible, though
feasibility should be judged by the laws of physics and not necessarily
current engineering limitations, and perform the tasks required by
a networking system.
We have already identified three primary tasks
that networking must accomplish: labeling and addressing,
determining paths in space and time for packets (routing and flow control),
and fragmentation/reassembly to match desired symbol size to actual glyph
capacity. We have also identified four basic physical properties
that are common to networking systems: space, time, energy, and distortion.
Three of these properties (space, time, and energy) are also of concern
to physically-situated applications, even in the absence of a networking
or storage system, and are candidates for general abstraction by
the operating system for {\em all} sensor/actuator applications.
Finally, we have described how generic computation, networking, and
storage systems may be described in terms of semantic layers that ultimately
reside within some physical artifact.
With these goals, tasks, and fundamental physical information properties
in hand we can now proceed to describe a layered meta-architecture in
greater detail. Figure~\ref{figure:system_meta_architecture} illustrates
our proposed meta-architecture as well as a decomposition of the
canonical network stack.
We will now briefly describe each layer of our meta-architecture.
We will not delve into many
interesting potential architectural details in this work, though
we are actively exploring such details and plan on reporting them
in future work.

\subsection{Meta-Architecture Layers}
\paragraph{The Glyph Layer}
The {\em Glyph} layer is aware and has some model of the actual
glyphs that will be used to transport symbols. An individual
glyph layer may be aware of only a single type of glyph, though
one could conceive of systems that provided ``fast path'' transitions
between different types of glyphs or combine multiple glyph types
into a single compound glyph. The Glyph layer is essentially a
sensor/actuator system and will likely encounter problems similar
to those found in sensor systems. This fact makes it seem that
access to the physical world should be controlled and consolidated
by the operating system at a level {\em below} the networking system
to avoid a ``stack inversion'' when a sensor application attempts
to use a networking system.
Space, time, energy, and distortion at the Glyph layer are likely
to be expressed in terms that are directly relevant to the 
glyph in question.
For example, a directional wireless system
has a ``native'' coordinate system based on its antenna pattern
and electromagnetic wave propagation.
Distortion may be expressed
directly in terms of relevant glyph attributes. Using Shannon bits
and signal processing provides an enormous boost in abstraction:
signal-to-noise ratios allow systems to model links
and perform power control in the context of control theory.

Routing and flow control at the Glyph layer have been extensively studied in
the context of MAC protocols, though most MAC layers do not perform multihop
routing. Flow control may coordinate with the Generic Symbol layer by
``backpressure'' and routes may use whatever time, energy, space, and
distortion metrics are sensible in a particular architecture.
Connecting to the generic symbol layer also involves
converting the transport glyph into a corresponding computational
glyph, and details of this conversion will vary depending on the system
in question. Using Shannon bits permits the formation of
an architectural waistline at the Glyph layer that is not apparent
from our generic meta-architectural diagram.

\paragraph{The Generic Symbol Layer}
The Generic Symbol Layer is where we expect primary architectural
waistlines to form, as is evidenced by the fanouts that exist
on either side of it. This layer is essentially an ``internetworking'' layer
that funnels in application requests for communication and fans them out
to Glyph layer devices for actual transport. The Generic Symbol layer
takes energy, time, and required address/label delivery constraints and
determines the best use of intermediate Glyph devices based on their
energy, time, distortion, and address/label delivery capabilities.
This layer will likely face some of the greatest computational
complexity as it must contend with potential combinatorial explosions
on either end. This layer also seems a likely candidate for division
into sublayers or some other more complicated systemic construct, though
it should be stressed that having a non-layered {\em internal} structure
of a layer in no way violates overall system layering.

\paragraph{The Application Symbol Layer}
The Application Symbol layer is where individual applications
may make networking and storage requests based on application-specific
requirements. These requirements are generally driven by clients
of the applications, and requests may be specified at a reasonably high level,
{\em e.g.} discovering labels and addresses of other computational entities
and exchanging information with them. Further layering of symbolic semantics
may occur within the Application Symbol layer, {\em e.g.} the little-used
OSI {\em Presentation} layer and the increasing use of XML as a standard
data interchange format.

\section{Conclusion}
In this work we have outlined a physically grounded way of thinking
about networking, computation, and storage and formulated a networking
system {\em meta-architecture} in this context. We believe that such a
meta-architecture is essential for organized analysis and discussion of
network architectures and systems, and are actively exploring details of
next-generation network architectures within the context of this
meta-architecture.

\bibliographystyle{unsrt}
\bibliography{physicalnetworking}

\end{document}